\documentclass[12pt]{article}
\pdfoutput=1
\usepackage{latexsym,cite}
\usepackage{amsmath}
\usepackage{amssymb}
\usepackage{amsthm}
\usepackage{subcaption}
\usepackage{mathtools}
\usepackage{braket}

\usepackage[top = 1. in,bottom = 1in, left = 1 in, right=1 in]{geometry}

\newcommand{\arXiv}[1]{\href{http://www.arXiv.org/abs/#1}{arXiv:#1}}
\newcommand{\doi}[2]{\href{http://dx.doi.org/#2}{#1}}
\usepackage[colorlinks=true, linkcolor=blue, bookmarks=true]{hyperref}

\allowdisplaybreaks

\makeatletter
\renewcommand\section{\@startsection {section}{1}{\z@}%
                  {-3.5ex \@plus -1ex \@minus -.2ex}
                  {2.3ex \@plus.2ex}%
                  {\normalfont\large\bfseries}}
\renewcommand\subsection{\@startsection{subsection}{2}{\z@}%
                   {-3.25ex\@plus -1ex \@minus -.2ex}%
                   {1.5ex \@plus .2ex}%
                   {\normalfont\bfseries}}
\makeatother

\newcommand{\beq}{\begin{equation}}
\newcommand{\eeq}{\end{equation}}

\newcommand{\sssty}{\scriptscriptstyle}
\newcommand{\ssty}{\scriptstyle}
\newcommand{\tsty}{\textstyle}
\newcommand{\dsty}{\displaystyle}
\newcommand{\ena}{\end{eqnarray}}
\newcommand{\beqa}{\begin{eqnarray}}
\newcommand{\eeqa}{\end{eqnarray}}
\newcommand{\bea}{\begin{eqnarray}}
\newcommand{\eea}{\end{eqnarray}}

\newcommand{\ad}{a^\dagger}
\newcommand{\la}{\lambda}
\newcommand{\Om}{\Omega}
\newcommand{\mn}{\mathrm{min}}
\newcommand{\mx}{\mathrm{max}}

\begin{document}
\begin{titlepage}
\begin{flushright}
\phantom{arXiv:yymm.nnnn}
\end{flushright}
\begin{center}
{\huge\bf Quantum estimates for\vspace{3mm}\\ classical polynomial optimization}
\vskip 12mm
{\large Oleg Evnin}
\vskip 3mm
{\it  High Energy Physics Research Unit, Faculty of Science, \\Chulalongkorn University,
Bangkok, Thailand\vspace{1mm} \\ Theoretische Natuurkunde, Vrije Universiteit Brussel (VUB) \\
\&\,\,
International Solvay Institutes, Brussels, Belgium}
\vskip 7mm
{\small {\tt oleg.evnin@gmail.com}}
\vskip 30mm
{\bf ABSTRACT}\vspace{3mm}
\end{center}
The problem of finding lower and upper bounds on multivariate homogeneous polynomials is both difficult and important given its applications to questions ranging from dynamical stability in complex potential landscapes to data analysis. From the standpoint of tensor eigenvalue theory, the question is equivalent to finding the smallest and the largest eigenvalues of the coefficient tensor corresponding to the given polynomial. Standard approaches outlined in the literature amount to running nonlinear iterations in search for the optimal rays along which the growth of the polynomial is fastest or slowest. Unlike the case of matrices (or their corresponding multivariate quadratic forms) convergence of such algorithms for higher-rank tensors is capricious due to the complex topography of polynomial objective functions. In this essay, a very different strategy, inspired by quantum-mechanical variational methods, is introduced for finding bounds on polynomials. The original polynomial is replaced by an operator acting in a suitably chosen (large) space of states, such that in an appropriate ``classical'' limit this operator approaches the original polynomial expression made of commutative variables. As a result, approximating the smallest and largest eigenvalues of the coefficient tensor, and thus finding bounds on polynomials, amounts to diagonalizing the resulting quantum operator, represented as a large matrix, and then inspecting the smallest and largest eigenvalues of this matrix. This approach is then successfully applied to standard test examples from tensor eigenvalue literature and other problems of interest in mathematical physics including Strichartz-type inequalities.\vspace{1cm}

\vfill

\end{titlepage}


\section{Introduction}

Consider a general homogeneous polynomial of an {\it even} degree $p$ in $d$ dimensions given by
\beq\label{Pdef}
P(\vec{x})\,\,\equiv \!\!\sum_{i_1,\cdots,i_p=1}^{d}\!\! C_{i_1\cdots i_p}x_{i_1}\cdots x_{i_p},
\eeq
where the rank $p$ {\it coefficient tensor} $C_{i_1\cdots i_p}$ is fully symmetric under index permutations. Is this polynomial positive-definite? Alternatively, can this polynomial be bounded from above by a simpler polynomial, for instance, something proportional to $(\sum_i x_i^2{\dsty)^{p/2}}$?

Such questions are known to be difficult. An obvious strategy is to look for a sum-of-squares decompositions, that is, a representation of $P$ as a sum of squares of degree $p/2$ polynomials. It is well-known since Hilbert that such decompositions do not necessarily exist even if $P$ is positive-definite, which paved the way to the formulation of Hilbert's 17th problem \cite{17th}.

In practical terms, since the value of $P$ at any point can be expressed through its values on the unit sphere $\sum_i x_i^2=1$ by homogeneity, to derive bounds on (\ref{Pdef}), one simply needs to find the global minimum and the global maximum of $P$ on the unit sphere. This is highly nontrivial, however, for higher degrees and dimensions, due to the well-known problem of complexity of landscapes of generic higher-dimensional functions \cite{complex1,complex2}, that is, the proliferation of critical points that obscure the search for global extrema.

The local extrema of $P$ on the unit sphere are determined, by the standard Lagrangian extremization, from
\beq\label{eigZ}
\sum_{i_2,\cdots,i_p=1}^{d}\!\! C_{i_1\cdots i_p}x_{i_2}\cdots x_{i_p}=\la x_{i_1},\qquad \sum_{i=1}^d x_i^2=1.
\eeq
This is familiar from the tensor eigenvalue theory \cite{tensortextbook} as one of the definitions of tensor eigenvalues, or more precisely, the definition of Z-eigenvalues in the terminology of Qi \cite{Qi}. Modern work on the tensor eigenvalue theory has been spurred by \cite{Qi,Lim}, while closely related concurrent consideration may be seen in \cite{DMtens} and all of these developments refer to the earlier treatment in \cite{GKZ} due to Gelfand, Kapranov and Zelevinsky. The counting of tensor eigenvalues has been developed starting with \cite{counting}, while modern studies of tensor eigenvalues from a mathematical physics perspective can be seen in \cite{randtens1,randtens2,randtens3}. Tensor eigenvalues find applications to a vast range of subjects from problems of nonlinear stability to data analysis topics, such as structural measures for hypergraphs \cite{benson}.

In the language of (\ref{eigZ}), the task of constructing bounds on $P$ can be expressed as finding the smallest and the largest among the eigenvalues $\la$, which can be denoted $\la_\mn$ and $\la_\mx$. If these {\it extreme eigenvalues} have been found, one evidently has the bounds
\beq\label{minmaxbound}
P(\vec{x})\ge \la_\mn \Big({\tsty\sum_{i=1}^d} x_i^2\Big)^{p/2},\qquad P(\vec{x})\le \la_\mx \Big({\tsty\sum_{i=1}^d} x_i^2\Big)^{p/2}.
\eeq
Searches for extreme eigenvalues of tensors have been discussed in the literature on numerical methods \cite{KR,KM,ADM,PSchr}. The main idea is to run (nonlinear) iterations whose fixed points include the extreme eigenvalues defined by (\ref{eigZ}). There are many more subtleties in this process than in the corresponding (linear) power iteration algorithms for matrix eigenvalue searches. The reason is in the dynamical complexity of nonlinear iterations that makes their convergence nontrivial and may lead to a local rather than a global extremum, even if the process converges at all. For that reason, one typically has to restart the iterations with many different initializations and compare the outputs.
In \cite{dynamical}, a continuous dynamical system is proposed as an alternative to discrete iterations, though the complex dynamical landscapes are still very much present. It is interesting to contemplate whether other strategies can be formulated that are more ``global'' in nature and thereby bypass the convergence subtleties.

The aim of this essay is to present an approach that reduces searches of extreme eigenvalues (\ref{eigZ}) and bounds on polynomials (\ref{Pdef}) to diagonalizing very large matrices, sidestepping all convergence issues of the iterative approaches. The most general idea is to replace the polynomial expression (\ref{Pdef}) with a quantum operator acting on some finite-dimensional space of states. In an appropriate ``classical'' limit, typically when the number of independent states becomes large, this operator should approach the original expression (\ref{Pdef}). An estimate for the extreme tensor eigenvalues corresponding to (\ref{Pdef}) is then obtained from the extreme eigenvalues of the large matrix representing the quantized operator. Such matrix diagonalization is completely straightforward, except that it would evidently become costly when the matrices are very large.

The origin of these ideas lies in the previous work of the author on resonant Hamiltonian systems \cite{quantres, resrev}. For these systems, whose Hamiltonians are homogeneous polynomials of the dynamical variables, it was observed that canonical quantization leads to extremely simple quantum operators that can be diagonalized numerically. By the usual classical-quantum correspondence, the minimal and maximal quantum energy eigenvalues must then approximate well the bounds on the classical Hamiltonian when the number of quantum particles becomes large and a semiclassical regime is approached. This idea has been tested in practice and novel, highly nontrivial polynomial inequalities were guessed in this manner \cite{qperiod1,qperiod2}, getting subsequently proved rigorously by mathematicians \cite{Schwinte}. In view of this success, it seems appropriate to formulate the method in the broadest possible terms and make it available for immediate use in other settings, which is the main rationale behind the current presentation.

The remaining technical exposition consists of two parts which are largely independent and can be read separately depending on the individual preferences.
The first part will discuss real-valued polynomials of complex variables whose coefficient tensors can be seen as a generalization of Hermitian matrices.
While such structures are not the most typical of polynomial optimization and tensor eigenvalue literature, they provide a setting that directly generalizes the previous
considerations for quantum resonant systems, and also connect naturally to the topic of bounds on Strichartz norms \cite{strnorm1,strnorm2,strnorm3,strnorm4} which has important applications in dynamics of nonlinear wave equations. The quantum operators utilized in the treatment of this problem will have the attractive form of conventional bosonic many-body Hamiltonians with few-body interactions. The second part will return to the more conventional and widely discussed formulation
for polynomials of real variables, as in (\ref{Pdef}), and describe quantum variational technology for constructing bounds of the form (\ref{minmaxbound}), followed by practical numerical implementation and discussion.

\section{Hermitian polynomials}

\subsection{Hermitian tensors, their polynomials and many-body Hamiltonians}

We will return in the next section to simple polynomials of real variables of the form (\ref{Pdef}), but for now, the focus will be on a slightly different starting point:
a degree $2q$ polynomial of $d$-dimensional complex variables $z^i$ given by
\beq\label{Hdef}
H(\vec{z})\,\,\equiv \!\!\sum_{i_1,\cdots,i_q=1}^d\sum_{j_1,\cdots,j_q=1}^d \!\! C_{i_1\cdots i_q,j_1\cdots j_q}\bar z_{i_1}\cdots \bar z_{i_q}z_{j_1}\cdots z_{j_q},
\eeq
with the coefficient tensor $C$ being fully symmetric under all permutations of the $i$-indices and $j$-indices among themselves and additionally satisfying
\beq
C_{i_1\cdots i_q,j_1\cdots j_q}=\overline{C_{j_1\cdots j_q,i_1\cdots i_q}}.
\eeq
The latter condition is a direct analog of Hermiticity for ordinary matrices and, for that reason, we will refer to $C$ as a {\it Hermitian tensor}. This terminology was used in \cite{htensor1,htensor2}, while tensors of this type also appear very naturally in considerations of interactions of weakly nonlinear waves \cite{resrev}. Note that $H(\vec{z})$ is real-valued.

Just like in the general outline presented in the introduction, deriving bounds on  $H(\vec{z})$ amounts to finding its absolute minimum and maximum on the unit sphere $\sum_i |z_i|^2=1$. Using a Lagrange multiplier, the local extremum condition is
\beq\label{Heig}
\sum_{i_2,\cdots,i_q=1}^d\sum_{j_1,\cdots,j_q=1}^d \!\! C_{i_1\cdots i_q,j_1\cdots j_q}\bar z_{i_2}\cdots \bar z_{i_q}z_{j_1}\cdots z_{j_q}=\la z_{i_1},
\qquad \sum_{i=1}^d |z_i|^2=1.
\eeq
The real numbers $\la$ satisfying this equation are referred to as {\it Hermitian eigenvalues}.

As explained in the introduction, the idea is to replace (\ref{Hdef}) with a quantum operator for which (\ref{Hdef}) is the classical limit. The operator can then be diagonalized as a matrix, and its extreme eigenvalues will provide estimates for the extreme values of $H$ on the unit sphere. For (\ref{Hdef}), a natural choice is provided by the quantum Hamiltonian
\beq\label{qH}
\mathcal{H}=\sum_{i_1,\cdots,i_q=1}^d\sum_{j_1,\cdots,j_q=1}^d \!\! C_{i_1\cdots i_q,j_1\cdots j_q}\ad_{i_1}\cdots  \ad_{i_q}a_{j_1}\cdots a_{j_q},
\eeq
where $a_i$ and $\ad_j$ are the standard bosonic creation-annihilation operators satisfying the commutation relations
\beq\label{comm}
[a_i,\ad_j]=\delta_{ij}.
\eeq
Indeed, for states localized in the phase space, such as coherent states, the expectation value of (\ref{qH}) will closely track the values of the original polynomial (\ref{Hdef}).
(Localization of quantum states in the phase space corresponding to $a_i$ has been discussed accurately in terms of Husimi densities in \cite{phasesp}.)

Physically, (\ref{qH}) is simply a particle-number-conserving Hamiltonian with $q$-body interactions for bosons with $d$ available modes, a construction completely common in condensed matter and many-body theory. As $\mathcal{H}$ commutes with the particle number operator
\beq\label{Ndef}
\mathcal{N}\equiv\sum_{k=1}^d \ad_k a_k,
\eeq
it can be diagonalized simultaneously with $\mathcal{N}$, that is, in a sector of states $|\Psi\rangle$ satisfying
\beq
\mathcal{N}|\Psi\rangle=N|\Psi\rangle.
\eeq 
As usual, $\mathcal{H}$ acts on the Fock space spanned by the vectors $|\eta_1,\cdots,\eta_d\rangle$ with nonnegative integer occupation numbers $\eta_1$, ..., $\eta_d$ given by
$$
|\eta_1,\cdots,\eta_d\rangle\equiv \left(\prod_{k=1}^d \frac{(a^\dagger_k)^{\eta_k}}{\sqrt{\eta_k!}}\right)|0,0,\cdots\rangle,\quad a_k |0,0,\cdots\rangle=0,\quad a_k^\dagger a_k|\eta_1,\cdots,\eta_d\rangle=\eta_k|\eta_1,\cdots,\eta_d\rangle.
$$
In the sector with $N$ particles, these occupation numbers satisfy $\sum_k\eta_k=N$. The total number of such states can be computed as the number of ways to place $d-1$ separators in $N+d-1$ positions, which gives the dimension of the space of states with $N$ particles:
\beq
D={N+d-1\choose d-1}.
\eeq

Restricted to the sector with $N$ particles, the Hamiltonian $\mathcal{H}$ thus becomes a $D\times D$ matrix. If we denote the smallest and largest eigenvalues of this matrix $E^{(N)}_\mn$ and $E^{(N)}_\mx$, a natural estimate for the extreme Hermitian eigenvalues is
\beq\label{conv}
\la_\mn=\lim_{N\to\infty}\frac{E^{(N)}_\mn}{\dsty N^q},\qquad \la_\mx=\lim_{N\to\infty}\frac{E^{(N)}_\mx}{\dsty N^q}.
\eeq
Indeed, consider the eigenvector $z^\mx_k$ corresponding to the maximal eigenvalue $\la_\mx$ in (\ref{Heig}), and construct the usual coherent state
\beq
|\Psi_\mx\rangle\sim \exp\left[s\sum_{k=1}^d z^\mx_k\ad_k\right]|0\rangle,
\eeq
with the appropriate normalization that we do not need to write explicitly, taking $s$ to be large. This state does not have definite values of either $\mathcal{N}$ or $\mathcal{H}$, but the uncertainties are small because $|\Psi_\mx\rangle$ is an eigenstate of $a_k$. Direct computation based on the commutation relations (\ref{comm}) will tell us that the expectation value of $\mathcal{N}$ is $s^2$, remembering that $\sum_k  |z^\mx_k|^2=1$, while its standard deviation is $O(s)$. Similarly, the expectation value of $\mathcal{H}$ is $s^{2q}\la_\mx$ while its standard deviation is $O(s^{2q-1})$. Both distributions are very narrowly concentrated at large $s$.
Expand then $|\Psi_\mx\rangle$ in terms of its projections $|\Psi^{(N)}_\mx\rangle$ on states with exactly $N$ particles:
$$
|\Psi_\mx\rangle=\sum_N|\Psi^{(N)}_\mx\rangle.
$$
By the above argument, this sum is strongly dominated by $N=s^2\pm O(s)$. This means that, within this dominant range, $\langle \Psi^{(N)}_\mx|\mathcal{H}|\Psi^{(N)}_\mx\rangle$ is itself within $O(s^{2q-1})$ of $s^{2q}\la_\mx$ while its variance is small, otherwise one would not be able to have a narrowly concentrated statistics of $\mathcal{H}$ in $|\Psi_\mx\rangle$ as it receives a large contribution from $|\Psi^{(N)}_\mx\rangle$. But this means that, the maximal eigenstate energy in the $N$-particle sector $E^{(N)}_\mx$ cannot be smaller than $s^{2q}\la_\mx$, up to the error $O(s^{2q-1})$, or $o(N^q)$, otherwise it would have in principle been impossible to make a combination of $N$-particle states with the expectation value of $\mathcal{H}$ in that range. In other words,
\beq\label{bound1}
N^{q}\la_\mx-E^{(N)}_\mx<o(N^q).
\eeq
Conversely, consider the state $|\Phi^{(N)}_\mx\rangle$ that is the actual top eigenstate of $\mathcal{H}$ in the $N$-particle sector:
\beq
\mathcal{H}|\Phi^{(N)}_\mx\rangle=E^{(N)}_\mx|\Phi^{(N)}_\mx\rangle,\qquad \mathcal{N}|\Phi^{(N)}_\mx\rangle=N|\Phi^{(N)}_\mx\rangle.
\eeq
This state can be expanded in terms of coherent states $|\vec{z},s\rangle\sim e^{s\sum_k z_k\ad_k}|0\rangle$, which lets one express $E^{(N)}_\mx$ as
\begin{align*}
&E^{(N)}_\mx=\frac{\langle\Phi^{(N)}_\mx|\mathcal{H}|\Phi^{(N)}_\mx\rangle}{\langle\Phi^{(N)}_\mx|\Phi^{(N)}_\mx\rangle} \\
&= \frac{{\dsty\int} [ds\, ds'\,d\vec{z}\,d\vec{z}^{\,'}]\langle\Phi^{(N)}_\mx|\vec{z},s\rangle\langle\vec{z}^{\,'},s'|\Phi^{(N)}_\mx\rangle\langle\vec{z},s|\vec{z}^{\,'},s'\rangle \,(ss'{\dsty )^{q}}\,\sum_{\{i,j\}}  C_{i_1...,j_1...}\bar z_{i_1}\cdots \bar z_{i_q}z'_{j_1}\cdots z'_{j_q}}{{\dsty \int} [ds\, ds'\,d\vec{z}\,d\vec{z}^{\,'}]\langle\Phi^{(N)}_\mx|\vec{z},s\rangle\langle\vec{z}^{\,'},s'|\Phi^{(N)}_\mx\rangle\langle\vec{z},s|\vec{z}^{\,'},s'\rangle },
\end{align*}
where $[ds\, ds'\,d\vec{z}\,d\vec{z}^{\,'}]$ is an appropriate hyperspherical integration measure for $\vec{z}$ and $\vec{z}^{\,'}$ residing on the unit sphere.
The integral is dominated by $s$ and $s'$ of order $\sqrt{N}$ since $|\Phi^{(N)}_\mx\rangle$ has a definite number of particles and $\langle\Phi^{(N)}_\mx|\vec{z},s\rangle$ decays rapidly away from those values. At the same time, $\langle\vec{z},s|\vec{z}^{\,'},s'\rangle$ is known explicitly as a Gaussian concentrated in the region $s|\vec{z}-\vec{z}^{\,'}|\sim 1$, and hence, up to errors that are $o(N^q)$ we can write
\beq
E^{(N)}_\mx\approx N^{q}\,\frac{{\dsty\int}\, [ds\,d\vec{z}]\,\,\,\big|\langle\vec{z},s|\Phi^{(N)}_\mx\rangle\big|^2 H(\vec{z})}{{\dsty\int}\, [ds\,d\vec{z}]\,\,\,\big|\langle\vec{z},s|\Phi^{(N)}_\mx\rangle\big|^2 }\le N^{q}\la_\mx.
\eeq
Hence, $E^{(N)}_\mx-N^{q}\la_\mx<o(N^{q})$, which together with (\ref{bound1} implies
\beq
\big|E^{(N)}_\mx-N^{q}\la_\mx\big|<o(N^{q}),
\eeq
and thus (\ref{conv}), given a similar argument for $\la_\mn$ and $E^{(N)}_\mn$.

Evidently, the above heuristic sketch may be improved substantially by more accurate estimates of the error terms, but it expresses physically nothing beyond the usual quantum-classical correspondence and the result is, in the sense, unsurprising. We will see below how the estimates of the form (\ref{conv}) work in practice.

\subsection{Resonant Hamiltonian systems and Strichartz norms}

Before proceeding with the first practical application of the ideas outlined above, it is wise to extend somewhat the definition of the objective function (\ref{Hdef}). There is a particular situation where the number of dimensions $d$ in (\ref{Hdef}) can be sent to infinity, while  the optimization approach advocated here remains completely tractable. This extension also connects to many interesting problems in mathematical physics.

Instead of (\ref{Hdef}), we will consider the following function
\beq\label{resdef}
H(\vec{z})\,\,\equiv \!\!\sum_{\substack{\sssty i_k,j_k=0\vspace{0.5mm}\\\sssty\sum_k i_k=\sum_k j_k}}^\infty \!\! C_{i_1\cdots i_q,j_1\cdots j_q}\bar z_{i_1}\cdots \bar z_{i_q}z_{j_1}\cdots z_{j_q}.
\eeq
Note the presence of the resonance condition $i_1+\cdots+i_q=j_1+\cdots+j_q$, while the number of variables is now infinite. Hamiltonian systems with Hamiltonians of the form (\ref{resdef}) emerge ubiquitously from approximations to the weakly nonlinear dynamics or nonrelativistic \cite{FGH,GHT, GT,BBCE,GGT,fennell,BEF} and relativistic \cite{FPU,CEV,BMR,CF,BEL,quintic} waves in resonant domains, as well as from the problem of composite operator renormalization in quantum field theory \cite{DM}. Strichartz norms of field variables, expanded in terms of the normal modes, can also be naturally recast in this form.

To construct bounds on (\ref{resdef}), we examine the corresponding quantum resonant Hamiltonian \cite{quantres}
\beq\label{qresdef}
\mathcal{H}\,\,\equiv \!\!\sum_{\substack{\sssty i_k,j_k=0\vspace{0.5mm}\\\sssty\sum_k i_k=\sum_k j_k}}^\infty \!\! C_{i_1\cdots i_q,j_1\cdots j_q}\ad_{i_1}\cdots \ad_{i_q}a_{j_1}\cdots a_{j_q}.
\eeq
There is more structure here than in the exposition for the finite-dimensional case (\ref{Hdef}) since $\mathcal{H}$ commutes with two operators:
\beq
\mathcal{N}\equiv\sum_{k=0}^\infty \ad_k a_k,\qquad \mathcal{M}\equiv\sum_{k=1}^\infty k\,\ad_k a_k.
\eeq
One can then diagonalize $\mathcal{H}$ simultaneously with $\mathcal{N}$ and $\mathcal{M}$, examine the eigenvalue patterns, and effectively conjecture on this basis inequalities relating (\ref{resdef}) and polynomial expressions in terms of the ``classical'' analogs of $\mathcal{N}$ and $\mathcal{M}$ given by
\beq\label{NMcl}
N_c\equiv \sum_{k=0}^\infty |z_k|^2,\qquad M_c\equiv\sum_{k=1}^\infty k\, |z_k|^2.
\eeq
For the sector of Fock vectors $|\eta_0,\eta_1,\cdots\rangle$ with
\beq
\mathcal{N}|\eta_0,\eta_1,\cdots\rangle=N|\eta_0,\eta_1,\cdots\rangle,\qquad \mathcal{M}|\eta_0,\eta_1,\cdots\rangle=M|\eta_0,\eta_1,\cdots\rangle,
\eeq
the occupation numbers evidently satisfy $\sum_k\eta_k=N$, $\sum_k k\,\eta_k=M$, and the number of solutions of these equations is given by $p_N(M)$, the number of integer partitions of $M$ into at most $N$ parts \cite{quantres}. Within such given $(N,M)$-sector, the Hamiltonian becomes a $p_N(M)\times p_N(M)$ matrix that is straightforwardly diagonalized. If we discover for the minimal and maximal Hamiltonian eigenvalues at given $N$ and $M$ that
\beq
E_\mn^{(N,M)}\ge F(N,M),\qquad E_\mx^{(N,M)}\le G(N,M)
\eeq
for some $F$ and $G$, it is natural to expect that (\ref{resdef}) satisfies the inequalities
\beq\label{resbounds}
H(\vec{z})\ge \lim_{t\to\infty}\frac{F(tN_c,tM_c)}{t^q},\qquad H(\vec{z})\le  \lim_{t\to\infty}\frac{G(tN_c,tM_c)}{t^q},
\eeq
with $N_c$ and $M_c$ given by (\ref{NMcl}). In this way, an effective heuristic routine is provided for guessing the correct inequalities that one can thereafter attempt to prove using conventional mathematical methods.

To demonstrate how this works in practice, consider the Lowest Landau Level (LLL) Hamiltonian given by
\beq
H_{LLL}\equiv \!\!\sum_{\substack{\ssty n,m,k,l=0\vspace{0.5mm}\\\ssty n+m=k+l}}^\infty \frac{(n+m)!}{2^{n+m+1}\sqrt{n!m!k!l!}}\,\,\bar z_n\bar z_m z_k z_l.
\eeq
This Hamiltonian has been studied extensively in relation to the physics of trapped Bose-Einstein condensates \cite{GHT,GT,BBCE,GGT,qLLL1,qLLL2,fetter}. One can easily see that $H_{LLL}$ is nonnegative by writing
\beq\label{LLL0}
H_{LLL}=\sum_{j=0}^\infty\sum_{k,l=0}^j  \frac{j!}{2^{j+1}\sqrt{k!(j-k)!l!(j-l)!}}\,\,\bar z_k\bar z_{j-k} z_l z_{j-l}=\sum_{j=0}^\infty  \frac{j!}{2^{j+1}}\Bigg|\sum_{k=0}^j\frac{z_k z_{j-k}}{\sqrt{k!(j-k)!}}\Bigg|^2,
\eeq
but is there a useful upper bound on $H_{LLL}$? By diagonalizing the corresponding quantum Hamiltonian $\mathcal{H}_{LLL}$ within individual $(N,M)$-sectors, one quickly finds out \cite{quantres} that the top eigenvalue is
\beq
E_{LLL,\mx}^{(N,M)}=\frac{N(N-1)}2.
\eeq
From (\ref{resbounds}) one then rightfully conjectures that
\beq\label{LLLupper}
H_{LLL}\le\frac{N_c^2}2,
\eeq
which can indeed be proved using a sum-of-squares decomposition \cite{shift}.

There are much less straightforward examples: thus, in \cite{FGH}, one finds a quartic Hermitian polynomial with a bigger set of modes (corresponding to the energy eigenstates of a two-dimensional harmonic potentials). This polynomial likewise satisfies, in our present notation, $H\le N_c^2/2$, which can be seen as a generalization of (\ref{LLLupper}), and the question of finding the correct coefficient 1/2 on the right-hand side is framed as finding ``sharp Strichartz estimates.'' From our present angle, the correct coefficients could be guessed effortlessly from inspecting the corresponding quantum eigenvalues. Further information on the relevant quantum setup and the upper bound problem can be found in \cite{shift}.

There are further lower bounds on $H_{LLL}$ beyond the trivial bound $H_{LLL}\ge 0$, and they have been actively discussed under the name of the ``yrast line'' in the literature related to the quantum Hall effect \cite{yrast,yrastrev}. One such inequality was deduced in \cite{qperiod1} precisely by inspecting the quantum eigenvalues of 
$\mathcal{H}_{LLL}$ and reads
\beq
H_{LLL}\,\ge\, \frac{N_c^2}2-\frac{N_cM_c-|Z|^2}{4},
\qquad Z \equiv\sum_{k=0}^\infty \sqrt{k+1}\, \bar z_{k+1}z_k,
\eeq
with $N_c\equiv\sum_k |z_k|^2$, $M_c\equiv \sum_k k |z_k|^2$.
A strategy was proposed for proving this nontrivial inequality that was subsequently brought to completion in \cite{Schwinte}. Further generalizations of such lower bounds to a large class of resonant Hamiltonians introduced in \cite{solvable} were developed in \cite{qperiod2}.

\subsection{An upper bound on a Hamiltonian of Biasi}\label{secFC}

The purpose of the presentation up to this point has been twofold: first, to describe in an abstract form applicable to various settings the technology for guessing inequalities for homogeneous polynomials used implicitly in \cite{qperiod1,qperiod2}; second, to briefly summarize various inequalities in the past literature where these methods were or could be useful. To demonstrate the power of the formalism, it remains to show how it can be employed to deduce a new nontrivial inequality that has not been reported before.

Consider the following class of resonant Hamiltonians introduced by Biasi \cite{Biasi} and explored further in \cite{BG,BCEJ}:
\beq\label{FC}
H_\sigma\equiv\frac12 \!\sum_{\substack{\ssty n,m,k,l=0\vspace{0.5mm}\\\ssty n+m=k+l}}^\infty\left[\frac{\sigma-1}2(n+m)+1\right] \frac{\sqrt{A^{(\sigma)}_nA^{(\sigma)}_mA^{(\sigma)}_kA^{(\sigma)}_l}}{A^{(\sigma)}_{n+m}}\,\,\bar z_n\bar z_m z_k z_l,
\eeq
where $\sigma\ge 1$ is a parameter, and $A_n^{(\sigma)}\equiv A_n(\sigma,1)$ with the standard Fuss-Catalan (or Raney) numbers, a combinatorial sequence appearing naturally in random matrix considerations \cite{Raney}:
\beq\label{FCnum}
A_n(\sigma,r)\equiv \frac{r\,\Gamma(\sigma n+r)}{\Gamma(n(\sigma-1)+r+1)\,\Gamma(n+1)}.
\eeq
As seen in \cite{Biasi,BG,BCEJ}, these Hamiltonians display remarkable dynamical properties including turbulent cascades that lead to formation of power-law spectra of $|z_n|^2$ in finite time and can be given an explicit analytic description. The precise, carefully engineered form of the coefficients is essential for these properties.

The Hamiltonian (\ref{FC}) is evidently positive, as in analogy with (\ref{LLL0}), one can rewrite it as
\beq\label{Hsbound}
H_\sigma=\frac12 \sum_{j=0}^\infty \,\frac{1+j(\sigma-1)/2}{A^{(\sigma)}_j}\,\, \Bigg|\sum_{k=0}^j \sqrt{A^{(\sigma)}_kA^{(\sigma)}_{j-k}}\,\,z_kz_{j-k}\,\Bigg|^2.
\eeq
But can one obtain a sensible upper bound? Given the complexity of the analytic expressions for the coefficients, the latter task is rather daunting in the absence of further inputs. And yet, with the technology described up to this point, one can easily quantize $H_\sigma$ and obtain its eigenvalues. An elementary Python script that performs this diagonalization is provided in Appendix~\ref{appFC}. One quickly discovers that, in each $(N,M)$-sector, the top eigenvalue changes linearly with $M$, and by fitting this dependence, extracts from the numerics\footnote{It is a legitimate question what uncertainties are involved in guessing analytic formulas from the results of numerical computations. In the concrete case we are considering, everything is straightforward, however. For integer values of $\sigma$, one observes that the top eigenvalue of the relevant quantum Hamiltonian within each $(N,M)$-sector is an integer up to tiny numerical errors. (One can easily run these computations with very modest computer resources for $N$ and $M$ up to 33.) Rounding off these tiny numerical errors then gives a working guess for the exact form of the top eigenvalues, which can then be tested further for a few noninteger values of $\sigma$.} the exact formula
\beq
E_{\sigma,\mx}^{(N,M)}=\frac{(N-1)(N+2\sigma M)}2.
\eeq
By (\ref{resbounds}), we then expect that $H_\sigma$ satisfies
\beq\label{FCineq}
H_\sigma\le \frac{N_c\,(N_c+2\sigma M_c)}2,
\eeq
with $N_c\equiv\sum_k |z_k|^2$, $M_c\equiv \sum_k k |z_k|^2$. There are no {\it a priori} guarantees that this new relation can be proved using sum-of-squares decompositions, but in this particular case, such a proof can indeed be constructed, and is given in Appendix~\ref{appFCproof}.

\section{Real polynomials}

For Hermitian polynomials of the form (\ref{Hdef}), one could design a many-body Hamiltonian (\ref{qH}) usable for constructing bound estimates.
This approach does not immediately transfer to polynomials of real-valued variables. For example, replacing a real-valued variable $x_i$ with the Hermitian operator $a_i+\ad_i$ will produce a quantum Hamiltonian that no longer conserves the particle number (\ref{Ndef}), so we will not be able to separate the sectors with different numbers of particles to obtain finite-dimensional matrices suitable for diagonalization. For that reason, returning to the polynomial (\ref{Pdef}) of real variables, a different strategy will have to be formulated.

\subsection{A variational problem for Z-eigenvalue estimates}

As explained in the introduction, finding the smallest and largest Z-eigenvalues $\la_\mn$ and $\la_\mx$, and hence constructing the bounds (\ref{minmaxbound})
amounts to identifying the global minimum and maximum of (\ref{Pdef}) on the unit sphere $\sum_i x_i^2=1$. One can parametrize this sphere by suitable angles that we will collectively denote $\Om$, and for each such set of angles there is a unit vector pointing in the corresponding direction $\vec{n}(\Omega)$ with $\sum_i n_i^2=1$.
We furthermore introduce the integration measure over the unit sphere normalized to 1:
\beq\label{intnorm}
\int d\Omega=1.
\eeq
With this, we can represent the minimum and maximum of $P(\vec{x})$ over the unit sphere as
\beq\label{laPsi}
\la_\mn=\inf_{\Psi(\Om)}\int d\Omega \,P(\vec{n}(\Omega))\,\Psi^2(\Om),\qquad \la_\mx=\sup_{\Psi(\Om)}\int d\Omega \,P(\vec{n}(\Omega))\,\Psi^2(\Om),
\eeq
where the minimization and maximization is over all real-valued functions $\Psi(\Om)$ satisfying
\beq\label{Psinrm}
\int d\Om \, \Psi^2(\Om)=1.
\eeq
This is now formulated\footnote{An anonymous referee has kindly pointed out that the mindset of this section has many parallels to the developments related to the Lasserre hierarchy \cite{Lasserre}, and in particular, its applications to polynomial optimization on the unit sphere \cite{dKL}. While the standard strategy in this latter approach reduces the original polynomial optimization to a semidefinite programming problem, there is a more recent variation relying on eigenvalue computations that appears closely related to the formulation reported here, see \cite{dKL,LJ}. The presentation in this essay provides an attractive physical angle on these prior considerations in the mathematical literature, as the relevant approximations to the exact polynomial optimization problem are interpreted naturally in terms of expectation values of quantum operators.} as a variational problem: if we extremize over all functions $\Psi(\Om)$, we will get the true extrema. To estimate the extrema, however, we can extremize instead over a particular, sufficiently large set of functions. One way to proceed is to extremize over all functions residing in a particular $D$-dimensional linear subspace:
\beq\label{Psie}
\Psi(\Om)=\sum_{I=1}^D \psi_I\, e_I(\Om)
\eeq
with a basis $e_I(\Om)$. Then, the condition (\ref{Psinrm}) turns into
\beq\label{unitM}
\sum_{I,J=1}^D M_{IJ}\psi_I \psi_J=1,\qquad M_{IJ}\equiv \int d\Om\,  e_I(\Om)\, e_J(\Om),
\eeq
and the extremization problems (\ref{laPsi}) turn into
\beq\label{laA}
\la_\mn\le\min_{\psi_I}\sum_{I,J=1}^D A_{IJ}\psi_I \psi_J,\qquad \la_\mx\ge\max_{\psi_I}\sum_{I,J=1}^D A_{IJ}\psi_I \psi_J,
\eeq
with
\beq
A_{IJ}\equiv\int d\Omega \,P(\vec{n}(\Omega))\, e_I(\Om)\, e_J(\Om).
\eeq
The local extrema of $\sum_{IJ} A_{IJ}\psi_I \psi_J$ subject to (\ref{unitM}) are now given by the (linear) generalized eigenvalue problem
\beq\label{AMeig}
\sum_{J=1}^D A_{IJ}\psi_J=\Lambda \sum_{J=1}^D M_{IJ}\psi_J,
\eeq
and the minimal and maximal matrix eigenvalues $\Lambda_\mn$ and $\Lambda_\mx$ provide practical estimates for the true global extrema $\la_\mn$ and $\la_\mx$, with convergence expected at large $D$.

The success of this procedure depends, of course, on the choice of the functional basis $e_I(\Om)$. Most generally, in order to compute the matrices $A$ and $M$, one will have to evaluate $O(D^2)$ multidimensional integrals. Done by brute force, say, numerically, that is likely to be impractical.

There is, however, a very natural and simple choice of the basis $e_I(\Om)$ that reduces the integrals defining $A$ and $M$ to analytically tractable expressions. Namely, we interpret $I$ as a collective index $\{i_1,i_2,\cdots,i_R\}$ for a fixed assignment of $R$, with $i_1\le i_2\le\cdots\le i_R$. The number of values such a collective index takes is 
\beq
D={R+d-1 \choose d-1}.
\eeq
We then define
\beq\label{eIdef}
e_I(\Om)\equiv \frac1{\sqrt{\mathcal{E}_I}}\,n_{i_1}(\Om)\cdots n_{i_R}(\Om),\qquad \mathcal{E}_I\equiv \int d\Om\, n^2_{i_1}(\Om)\cdots n^2_{i_R}(\Om).
\eeq
With this choice, $A$ and $M$ are expressed in terms of the well-known integrals of monomials of the unit vector components \cite{sphere}
\beq\label{nint}
\int d\Om\,\, n_1^{2a_1}(\Om)\,n_2^{2a_2}(\Om)\cdots n_d^{2a_d}(\Om)=\frac{(d-2)!!}{(d-2+2\sum_k a_k)!!}\prod_{\substack{\ssty k=1\vspace{0.5mm}\\\ssty a_k\ne 0}}^d (2a_k-1)!!\,.
\eeq
(If, on the other hand, any of the powers is not an even number, the integral vanishes due to reflection symmetry.) One way to derive this formula is to observe that, due to rotation and reflection symmetries, $\int d\Om\, n_{i_1}\cdots n_{i_S}$ vanishes if $S$ is odd and should otherwise be proportional to a fully symmetrized product of Kronecker symbols $\delta_{\{i_1i_2}\delta_{i_3i_4}\cdots\delta_{i_{S-1}i_S\}}$. This, in turn, can be expressed as a sum over all distinct partitions of the index set into $S/2$ pairs. The proportionality coefficient can be fixed recursively by observing that contracting $i_{S-1}$ with $i_S$ gives the same integral with $S\to S-2$ until one reaches 
(\ref{intnorm}) with $S=0$. Finally one can compute the number of matching pairings in an index set $(i_1,i_2,\cdots,i_S)$ where 1 occurs $2a_1$ times, 2 occurs $2a_2$ times, etc, yielding (\ref{nint}). Note that the basis (\ref{eIdef}) is not orthonormal and the normalization factors $\mathcal{E}_I$ are introduced to make the matrix $M$ better conditioned.

With these specifications and a given polynomial $P$, the computation of $A$ and $M$ is algorithmic, and the question is how the estimates for the extreme tensor eigenvalues $\la_\mn$ and $\la_\mx$ obtained from the generalized eigenvalue problem (\ref{AMeig}) vary as $R$ increases.  Note that a similar strategy could in principle be applied to Hermitian polynomials as an alternative to the many-body Hamiltonians we utilized in the previous section, since a Hermitian polynomial can be expressed in terms of $\mathrm{Re}\,\vec{z}$ and $\mathrm{Im}\,\vec{z}$.

The decomposition (\ref{Psie}) with the basis (\ref{eIdef}) can be interpreted as a truncated hyperspherical harmonic expansion. Indeed, the coefficients $\psi_I\equiv \psi_{i_1\cdots i_R}$ can be viewed as the components of a fully symmetric rank $R$ tensor. Expanding this tensor into the traceless parts and traces of various lower orders yields one of the standard representations of the hyperspherical harmonic expansion up to rank $R$ in $d$ dimensions, as dictated by the representation theory of the rotational group $SO(d)$. Truncated spherical harmonic expansions often appear in relation to ``fuzzy spheres,'' a simple form of noncommutative spaces \cite{fuzzy} and, in this sense, noncommutativity is implicit in the quantum variational problem (\ref{laPsi}).

\subsection{Examples}

The procedure outlined above can be straightforwardly implemented numerically, and a simple script is provided in Appendix~\ref{appZ}. This produces the following results for a few concrete polynomials used as test cases in the tensor eigenvalue literature:

\subsubsection*{\it The Motzkin polynomial}

A classic polynomial mentioned in discussions of positivity and sum-of-squares decomposition is the following expression due to Motzkin:
$$
P_{Motzkin}(x,y,z)=x^4y^2+x^2y^4+z^6-3\,x^2y^2z^2.
$$
This polynomial is nonnegative by virtue of the arithmetic-geometric mean inequality $a+b+c\ge 3\,\sqrt[3]{abc}$ applied to $a=x^4y^2$, $b=x^2y^4$, $c=z^6$, and yet it cannot be represented as a sum of squares of degree 3 polynomials. The maxima on the unit sphere $x^2+y^2+z^2=1$ are attained at $z=\pm 1$ where $P_{Motzkin}=1$. The coefficient tensor of this polynomial appears in Example 6 of \cite{ADM}.

Applying the general technology outlined above and the script from Appendix~\ref{appZ} produces rapid convergence to the correct extreme eigenvalues $\la_\mn=0$ and $\la_\mx=1$. Thus, at $R=20$ with $D=231$, one gets the estimates 0.003754 and 0.96769 for $\la_\mn$ and $\la_\mx$. At $R=30$ with  $D=496$, this improves to 0.001938 and 0.984188, and at $R=45$ with $D=1081$, to 0.000955 and 0.992498. The error appears to decay as roughly $1/D$.

\subsubsection*{\it Example 1 from  \cite{ADM}}

The following rank 4 tensor in $d=3$ dimensions is often evoked as a test example in discussions of numerical methods for tensor eigenvalues, going back to \cite{KR}:
\begin{align*}
&a_{1111}=0.2883,\quad a_{1112}=-0.0031,\quad a_{1113}=0.1973,\quad a_{1122}=-0.2485,\quad a_{1123}=-0.2939,\\
&a_{1133}=0.3847,\quad a_{1222}=0.2972,\quad a_{1223}=0.1862,\quad a_{1233}=0.0919,\quad a_{1333}=-0.3619,\\
&a_{2222}=0.1241,\quad a_{2223}=-0.3420,\quad a_{2233}=0.2127,\quad a_{2333}=0.2727,\quad a_{3333}=-0.3054. 
\end{align*}
Its $\la_\mn$ and $\la_\mx$ are known as -1.0954 and 0.8893 respectively, For these, the script from Appendix~\ref{appZ} produces the following estimates: at $R=20$  with $D=231$, -1.0456657 and 0.8603242; at $R=30$ with  $D=496$, -1.071 and 0.875; at $R=45$ with $D=1081$, -1.08383 and 0.88255. The error decays roughly as $1/D$.

\subsubsection*{\it Example 2 from \cite{ADM}}

In \cite{ADM}, the following rank 4 tensor in $d=5$ dimensions is evoked
$$
a_{ijkl}=\sin(i+j+k+l),\qquad 1\le i,j,k,l\le 5.
$$
Its $\la_\mn$ and $\la_\mx$ are -8.8463 and 7.2595. For these, the script from Appendix~\ref{appZ} produces the following estimates: at $R=15$ with $D=3876$, -7.82 and 6.36; at $R=20$ with $D=10626$,  -8.19365 and 6.6865. This suggests an error decaying as roughly $1/\sqrt{D}$. It would be interesting to get a more complete picture of the $D$-dependence of the error based on the convergence analysis in the style of \cite{dKL}.

\section{Discussion}

Standard methods for constructing bounds on homogeneous multivariate polynomials, or equivalently, for finding the extreme eigenvalues of their coefficient tensors,
revolve around running nonlinear iterations and hence their convergence is precarious due to the complex topography of higher-degree multivariate polynomial functions.
The complementary approach advocated here relies on completely different principles and aims for replacing the original polynomial with a quantum operator that is then diagonalized using standard spectral methods. The trade-off is between the need to diagonalize potentially extremely large matrices, as is common in quantum problems, and having immediate global access to the entire configuration space, sidestepping all issues caused by local extrema and other topographic peculiarities. This approach has been demonstrated to be applicable for Hermitian polynomials of the form (\ref{Hdef}), their infinite-dimensional generalizations (\ref{resdef}) and ordinary multivariate polynomials of real variables (\ref{Pdef}). As a result of making use of such heuristics, a novel inequality (\ref{FCineq}) was deduced for the highly nontrivial polynomial (\ref{FC}), while adequate performance has also been demonstrated for finding the extreme eigenvalues of a few standard small tensors used as test examples in the  literature on numerical methods.

Many possible improvements and extensions can be envisaged and, indeed, the numerical implementation provided in the appendices is very basic and serves predominantly demonstrative purposes. One could naturally try to combine the approach developed here for a sort of crude global optimization followed by local optimization based on conventional iterative methods. The focus of the exposition here has been on recovering bounds on polynomials, but not on the independent variable configurations that saturate (or approximate) these bounds. The latter question could naturally be approached by examining the expectation values $\langle \ad_n a_m\rangle$ and $\langle n_i n_j\rangle$ evaluated on the extremizer wavevectors arising from the constructions presented here. It is known, however, that the convergence of wavevectors in quantum variational methods is much less robust than the convergence of the eigenvalues, and depends, in particular, on spectral gaps. In the context of polynomial optimization, one may expect that the difficulty of recovering the extremizer configurations depends on whether the bottom and the top of the objective function landscape are relatively flat.

There are two main aspects of the potential applications for the technology described here. First, it can serve as a heuristic aid for analytic work, as in \cite{qperiod1, qperiod2} or in section~\ref{secFC}. In this situation, one simply uses the quantum estimates as a way to guess the correct coefficients in the inequalities for the objective function that one would then hope to prove analytically. Evidently, knowing which precise inequality one should be proving is of great help for guiding any analytic work. Intuitively, one should expect that the polynomials arising from such models are defined by explicit, relatively simple analytic expressions, their topography is relatively tame, and rather modest numerical resources will suffice for obtaining a good approximation to the correct bounds. Once conjectural inequalities have been inferred from this sort of quantum heuristics, the prospects of proving them using sum-of-squares decompositions can be further assessed by running numerical sum-of-squares tools based on semidefinite programming \cite{SDP}.

On the opposite end are the applications to data analysis, as in finding the eigenvalues of large tensors that are effectively random. In this case, one expects to
confront the complexity of multivariate nonlinear landscapes head-on. This will likely lead to the need to consider extremely large matrices in the auxiliary quantum problems. Luckily, what interests us the most are the extreme eigenvalues, which are usually recovered much more easily than the full spectrum. The Lanczos algorithm is, in particular, often mentioned as an effective approach to very large matrices that is especially suitable for recovering the extreme eigenvalues. Finally, all the problems related to large matrices are bypassed if one gets access to a quantum computer, where spaces of states with dimensions of order $10^{100}$ or more would be easily attainable with moderately-sized hardware. The fact that polynomial optimization problems have been reformulated here in the language of quantum mechanics makes them naturally adapted for such treatments.

\section*{Acknowledgments}

I am indebted to Marine De Clerck for collaboration on closely related subjects and multiple important discussions.

\appendix

\section{A Python script for deducing the bound on Biasi's Hamiltonian}\label{appFC}

\begin{verbatim}
import math
import numpy as np
import scipy.special as sp

N=15
M=30
s=9.73

# generate all partitions of M into at most N parts as sets of occupation 
# numbers, in a list of vectors
def part (N,M):
 if M==0:
  return [[N]]
 else:
  if N==1:
   return [[0 for i in range(M)]+[1]]
  else:
   prt=[[i[0]+1]+i[1:] for i in part(N-1,M)]
   if M>=N:
    prt=prt+[[0]+i+[0 for k in range(0,N-1)] for i in part(N,M-N)]
   return prt

#act with H on a vector of the form 
#[[component,[list of occupation numbers]], [comp...,[]],...]
def applyH (vec):
 vecout=[]
 for j in range(M+1):
  for k in range(int(math.floor(j/2))+1):
   for l in range(int(math.floor(j/2))+1):
    for c in range(len(vec)):
     outcomp=[]
     if l!=j-l:
      if (vec[c][1][l]>0) and (vec[c][1][j-l]>0):
       outcomp=vec[c][1][:]
       outcoeff=vec[c][0]
       outcoeff=2*outcoeff*Cjkl[j][k][l]*sq[outcomp[l]]*sq[outcomp[j-l]]
       outcomp[l]=outcomp[l]-1
       outcomp[j-l]=outcomp[j-l]-1
       if k!=j-k:
        outcoeff=2*outcoeff*sq[outcomp[k]+1]*sq[outcomp[j-k]+1]
        outcomp[k]=outcomp[k]+1
        outcomp[j-k]=outcomp[j-k]+1
       else:
        outcoeff=outcoeff*sq[outcomp[k]+1]*sq[outcomp[k]+2]
        outcomp[k]=outcomp[k]+2
     else:
      if (vec[c][1][l]>1):
       outcomp=vec[c][1][:]
       outcoeff=vec[c][0]
       outcoeff=outcoeff*Cjkl[j][k][l]*sq[outcomp[l]]*sq[outcomp[l]-1]
       outcomp[l]=outcomp[l]-2
       if k!=j-k:
        outcoeff=2*outcoeff*sq[outcomp[k]+1]*sq[outcomp[j-k]+1]
        outcomp[k]=outcomp[k]+1
        outcomp[j-k]=outcomp[j-k]+1
       else:
        outcoeff=outcoeff*sq[outcomp[k]+1]*sq[outcomp[k]+2]
        outcomp[k]=outcomp[k]+2
     if len(outcomp)>0:
      vecout.append([outcoeff,outcomp])
 return vecout

#sort the components of a vector according to the order 
#induced by the partition generator
def sortvec (vec):
 vecout=[[0,partlist[i]] for i in range(len(partlist))]
 for i in range(len(partlist)):
  for c in range(len(vec)):
   if partlist[i]==vec[c][1]:
    vecout[i][0]=vecout[i][0]+vec[c][0]
 return vecout  

partlist=part(N,M)
dim=len(partlist)

sq=[i**0.5 for i in range(M+1)]

#Biasi's Fuss-Catalan interaction coefficients
FC=[sp.binom(s*j+1,j)/(s*j+1) for j in range(M+1)]
sqFC=[FC[j]**0.5 for j in range(M+1)]
Cjkl=[[[(1+(s-1)*j/2)*sqFC[k]*sqFC[j-k]*sqFC[l]*sqFC[j-l]/(2*FC[j]) for l in\
 range(int(math.floor(j/2))+1)] for k in range(int(math.floor(j/2))+1)]\
 for j in range(M+1)]

A=[]
for i in range(dim):
 vec=sortvec(applyH([[1,partlist[i]]]))
 A=A+[[vec[j][0] for j in range(dim)]]

w=np.linalg.eigvalsh(A)
print(w[0])
print(w[-1],(N-1)*(N+2*s*M)/2)
\end{verbatim}

\section{A proof of the upper bound on Biasi's Hamiltonian}\label{appFCproof}

As we set out to prove the inequality (\ref{FCineq}) guessed using the quantum heuristics developed here, practical guidance is provided by the special case $\sigma=1$ which leads to considerable simplifications as $A_n^{(1)}\equiv 1$. This case was introduced by G\'erard and Grellier \cite{GG} and studied in extensive subsequent work as it corresponds to a remarkable integrable Hamiltonian with distinctive dynamical properties. The bound (\ref{FCineq}) is easily proved in this case, see \cite{superQFT} for a compact presentation compatible with the notation here. The structure of this proof provides hints for deriving (\ref{FCineq}), though the complicated form of the coefficients based on the Fuss-Catalan numbers at general $\sigma$ makes the derivation algebraically more intricate.

In practice, consider the following transformation of the right-hand side of (\ref{FCineq}):
\beq
 \frac{N_c\,(N_c+2\sigma M_c)}2\equiv\frac12 \sum_{k,l=0}^\infty (1+2k\sigma)|z_k|^2|z_l|^2=\frac12 \sum_{j=0}^\infty\sum_{k=0}^j (1+j\sigma)\,|z_k|^2|z_{j-k}|^2.
\eeq
We can then write
\beq
1+j\sigma=2\,\frac{\Gamma(2+j(\sigma-1))}{\Gamma(3+j(\sigma-1))}\frac{\Gamma(2+j\sigma)}{\Gamma(1+j\sigma)}\left[1+j(\sigma-1)/2\right]=\frac{A_j(\sigma,2)}{A_j(\sigma,1)}\left[1+j(\sigma-1)/2\right],
\eeq
with the definition (\ref{FCnum}). A standard convolution property of the Fuss-Catalan numbers reads
\beq
A_j(\sigma,r+r')=\sum_{l=0}^{j} A_l(\sigma,r)A_{j-l}(\sigma,r').
\eeq
From this, with $r=r'=1$, one gets
\beq
 \frac{N_c\,(N_c+2\sigma M_c)}2=\frac12 \sum_{j=0}^\infty\frac{1+j(\sigma-1)/2}{A_j^{(\sigma)}}\sum_{k,l=0}^j |z_k|^2|z_{j-k}|^2 A_l^{(\sigma)}A_{j-l}^{(\sigma)}.
\eeq
Then, from (\ref{Hsbound}),
$$
 \frac{N_c\,(N_c+2\sigma M_c)}2-H_\sigma=\frac14 \sum_{j=0}^\infty\frac{1+j(\sigma-1)/2}{A_j^{(\sigma)}}\sum_{k,l=0}^j \Bigg|\sqrt{A_k^{(\sigma)}A_{j-k}^{(\sigma)}}\,z_lz_{j-l}-\sqrt{A_l^{(\sigma)}A_{j-l}^{(\sigma)}}\,z_kz_{j-k}\Bigg|^2,
$$
which is manifestly nonnegative.

\section{A Python script for extreme Z-eigenvalues}\label{appZ}

\begin{verbatim}
import numpy as np
import scipy.linalg as LA

r=20

#Motzkin polynomial
d=3
deg=6
poly=[[1.0,[1,1,1,1,2,2]],[1.0,[1,1,2,2,2,2]],[-3.0,[0,0,1,1,2,2]],\
 [1.0,[0,0,0,0,0,0]]]

#E. Kofidis and P. A. Regalia, SIAM J Matri Anal App 23 (2002) 863
#d=3
#deg=4
#poly=[[0.2883,[0,0,0,0]],[-0.0031*4,[0,0,0,1]], [0.1973*4,[0,0,0,2]],\
# [-0.2485*6,[0,0,1,1]],  [-0.2939*12,[0,0,1,2]],[0.3847*6,[0,0,2,2]],\
# [0.2972*4,[0,1,1,1]], [0.1862*12,[0,1,1,2]], [0.0919*12,[0,1,2,2]],\
# [-0.3619*4,[0,2,2,2]], [0.1241,[1,1,1,1]], [-0.3420*4,[1,1,1,2]],\
# [0.2127*6,[1,1,2,2]], [0.2727*4,[1,2,2,2]], [-0.3054,[2,2,2,2]]]

#degree 4 with a formula for the coefficient tensor
'''d=5
deg=4
fac=[1]
for i in range(1,d+1):
 fac+=[fac[-1]*i]
poly=[]
for i in range(d):
 for j in range(i,d):
  for k in range(j,d):
   for l in range(k,d):
    count=np.zeros((d))
    count[i]+=1
    count[j]+=1
    count[k]+=1
    count[l]+=1
#Example 2 from Computing tensor Z-eigenpairs via an alternating direction method
    poly+=[[(fac[deg]/np.prod([fac[int(count[s])] for s in range(d)]))\
      *np.sin(i+j+k+l+4),[i,j,k,l]]]'''

dfac=[1.0]
for i in range(1,r+deg//2+1):
 dfac+=[dfac[-1]*(2*i-1)]

#generate sets of num entries from elements in ascending order
def genincr (num,elements):
 if num==1:
  return [[element] for element in elements]
 else:
  list=[]
  for i in range(len(elements)):
   tails=genincr(num-1,elements[i:])
   for tail in tails:
    list+=[[elements[i]]+tail]
  return list

#count matching complete pairings among elements
def countpairings(elements):
 count=1
 for dir in range(d):
  howmany=len(np.nonzero(np.array(elements)==dir)[0])
  if howmany!=0:
   if howmany%2!=0:
    return 0
   else:
    count*=dfac[howmany//2]
 return count

vecs=genincr(r,range(d))
dim=len(vecs)

M=np.zeros((dim,dim))
nrmM=np.prod([(d+2*i)*1.0 for i in range(r)])
sqnrm=np.zeros((dim))

for i in range(dim):
 sqnrm[i]=np.sqrt(countpairings(vecs[i]+vecs[i])/nrmM)

for i1 in range(dim):
 for i2 in range(i1,dim):
  M[i1,i2]=countpairings(vecs[i1]+vecs[i2])/(sqnrm[i1]*sqnrm[i2]*nrmM)
  if i2>i1:
   M[i2,i1]=M[i1,i2]

A=np.zeros((dim,dim))
nrmA=np.prod([(d+2*i)*1.0 for i in range(r+deg//2)])

for i1 in range(dim):
 print(i1,'/',dim)
 for i2 in range(i1,dim):
  entry=0
  for term in poly:
   entry+=term[0]*countpairings(vecs[i1]+term[1]+vecs[i2])
  A[i1,i2]=entry/(sqnrm[i1]*sqnrm[i2]*nrmA)
  if i2>i1:
   A[i2,i1]=A[i1,i2]

eigs,eigvec=LA.eigh(A,b=M)

print(eigs[0])
print(eigs[-1])
\end{verbatim}

\end{document}